# EFFECT OF LASER INTENSITY AND DYNAMICS OF PLASMA ON LASER INDUCED BREAKDOWN SPECTROSCOPY

V. N. Rai[1] and Jagdish P. Singh[2, 3]

[1]Raja Ramanna Centre for Advanced Technology

Indore-452013 (INDIA)

vnrai@rrcat.gov.in

[2]Institute for Clean Energy Technology

Mississippi State University

205 Research Boulevard

Starkville, MS 39759 (USA)

[3]JPS Advanced Technology LLC

202 Setter Street

Starkville, MS 39759 (USA)

## ABSTRACT

Laser-induced breakdown spectroscopy (LIBS) show enhancement in its signal, when the laser-induced plasma is confined/decelerated under the effect of an external steady magnetic field or in a small cavity. An enhancement in LIBS signal has been observed ~2 times in the case of magnetic confinement. Combination of magnetic and spatial confinement provide enhancement by an order of magnitude. Theoretical analysis of the decelerated plasma has been found in agreement with the experimental observations. The enhancement in LIBS signal is found dependent on the efficiency of plasma confinement as well as on the time duration of laser. The saturation in LIBS signal at higher laser intensity is found correlated with electron-ion collision frequency as well as on the dynamics and instability of plasma plume. Possibility of further enhancement in emission has also been discussed.

## 1. INTRODUCTION

Various types of phenomena take place during laser matter interaction such as creation of plasma, material ablation, emission of radiation ranging from visible to X-ray wavelength, generation of high energy electrons and ions as well as different waves and instabilities [1-2]. Dynamics of laser produced plasma play an important role in deciding the characteristics of the plasma [1-8]. Properties of laser produced plasma mainly depend on the characteristics of the laser being used for producing the plasma. The laser produced plasma expands normally to the target surface and remains independent on the angle of incidence of laser on the target material. The free expansion of plasma depends mainly on the density and temperature of the plasma, which ultimately depends on laser intensity and its time duration. Any change in the free expansion of plasma affects all of its associated properties. The emission from laser-produced plasma under different conditions is an important subject of investigation in many laboratories due to its technological application in various fields of research such as material science, chemical physics, plasma physics as well as inertial and magnetic confinement fusion [1-2, 9-15]. Optical emission spectroscopy of the laser-produced plasma has been investigated by many laboratories in the last two decades in the name of laser-induced plasma (LIP) emission as well as laser-induced breakdown spectroscopy (LIBS). LIBS has been established as an analytical tool for in-situ determination of the elemental composition of the target samples in any form (solid, liquid, gas and aerosols) with fast response and high sensitivity without any sample preparation and surface treatment[16]. Along with the wide applicability of LIBS in various fields, it has also been used for the characterization of laser-induced plasma processes occurring during the production of thin solid films and nanoparticles of different materials. These aspects have been studied theoretically and the related models have been applied successfully to interpret the data obtained in a wide range of experimental conditions. The theoretical study of emission during LIBS has been found useful in better understanding the physical processes taking place in the plasma, which is beneficial in many ways particularly in getting optimum experimental parameters for this technique along with enhancing the sensitivity of this analytical system [17-21].

Various techniques have been reported to increase the plasma emission (LIBS signal), which decides the emission intensity and consequently the sensitivity of the LIBS [15-16]. Mainly magnetic and spatial confinement of the laser produced plasma as well as dual pulse excitation is important techniques, which enhances the intensity of LIBS by an order of magnitude [22-25]. It is supposed that kinetic energy of the plasma is transformed into thermal energy during the plasma confinement (magnetic or spatial), which helps in heating and exciting the atomic species of the plasma resulting in enhanced plasma emission. In double pulse excitation, second laser pulse gets absorbed in the plasma created by the first laser pulse resulting in enhanced plasma emission. In this case enhancement in emission is mainly due to heating of the plasma by the second laser pulse as well as by increased ablation of material from the target. Recently a combination of magnetic and spatial confinement has been used effectively to improve the emission from the plasma manifold (24 times) [25]. In this case, the plasma is confined by a magnetic field as well as by the reflected shock wave from the walls of a small cavity. It has been reported that the combination of both the techniques compresses the plasma at the center of cavity resulting in an increased rate of collisions among the plasma particles, which leads to an increase in the number of atoms in higher energy state followed by enhanced emission intensity. However the saturation of atomic emission from the plasma inherently limits the laser intensity coupled to plasma that is useful for optimizing/enhancing the LIBS sensitivity. Saturation in plasma emission with an increase in laser intensity occurs comparatively at lower laser intensity in the presence of magnetic confinement [26]. The study of saturation process and its dependence on different plasma parameters has been performed in order to enhance the efficiency of plasma emission [21]. The theoretical basis for enhancement in plasma emission as a result of magnetic and spatial confinement as well as under double pulse excitation has been reported earlier [20-21]. Still very little theoretical work is known about the process of plasma emission and saturation under the effect of variation in plasma parameters and confinement.

This paper presents a brief reviews to better understand the effect of laser intensity and the dynamics of laser produced plasma during LIBS under different experimental conditions. An effect of time duration of laser on LIBS performance has also been discussed.

## 2. EMISSION FROM LASER-INDUCED PLASMA IN MAGNETIC FIELD

The expansion and emission of the plasma in the presence of a nearly uniform magnetic field can be understood in a simple manner by equating the initial kinetic energy of the plasma with that of magnetic energy in the volume $(4/3)\pi R_B^3$, where $R_B$ is defined as the magnetic confinement (plasma bouncing) radius [1,11]. This can be written as

$$\frac{1}{2}Mv_0^2 = \left(\frac{B_0^2}{8\pi}\right) \bullet \left(\frac{4}{3}\pi R_B^3\right) \quad (1)$$

Where M is the mass of the plasma, $v_0$ is the initial plasma expansion velocity, and $B_0$ is the ambient magnetic field. From Eq.-1, it is possible to obtain the bouncing radius of the plasma as

$$R_B = \left(\frac{3Mv_0^2}{B_0^2}\right)^{\frac{1}{3}} \quad (2)$$

This shows that plasma expansion will be fully stopped near the bouncing radius $R_B$, as a result of plasma deceleration in the presence of the magnetic field, if the plasma is perfectly conducting. The deceleration of the plasma expansion under the effect of magnetic field can be given as [8]

$$\frac{v_2}{v_1} = \left(1 - \frac{1}{\beta}\right)^{\frac{1}{2}} \quad (3)$$

where $v_1$ and $v_2$ are the plasma expansion velocities in the absence and the presence of magnetic field. The plasma β ($=\frac{8\pi nkT_e}{B^2}$) is a ratio of kinetic energy of plasma to the magnetic energy. Eq.-3 clearly indicates that the plasma confinement (deceleration) will be effective only when the plasma β is low, that is, either the plasma temperature and density (or both) are low or the magnetic field becomes too high.

It is possible to write the ratio of plasma emission in the presence ($I_2$) and the absence ($I_1$) of the magnetic field as [23]

$$\frac{I_2}{I_1} = \left[\frac{v_1 t_1}{v_2 t_2}\right]^3 \quad (4)$$

where $t_1$ and $t_2$ is the plasma emission time in the absence and presence of magnetic field. Eq. 4 indicates that plasma emission intensity will be inversely proportional to the cube of the size (product of expansion velocity and the emission time) of the plasma plume. Eq. - 3 defines the ratio of plasma expansion velocity in term of plasmaβ, which can be used in Eq.-4. Finally, the ratio of plasma emission in the presence and absence of magnetic field can be written as

$$\frac{I_2}{I_1} = \left(1 - \frac{1}{\beta}\right)^{-\frac{3}{2}} \left(\frac{t_1}{t_2}\right)^3 \quad (5)$$

This expression indicates that the ratio of plasma emission in the presence and absence of magnetic field is mainly dependent on plasma β as well as on the ratio of time duration of emission from the plasma. According to eq.-3 and 5 plasma deceleration and enhancement in emission in the presence of magnetic field is possible only for low β plasma, that is, when either plasma temperature and density or both become low after the plasma expansion.

## 3. EMISSION FROM DOUBLE AND SINGLE LASER PULSE EXCITATION

### 3.1 Emission from Double Pulse LIBS

First emission from double pulse LIBS will be discussed here, which will be simplified to single pulse LIBS. A simplified expression for enhancement in emission during double pulse excitation has been reported earlier [20]. In this case, it has been considered that second laser pulse gets absorbed in the plasma produced by first laser pulse after its expansion for the time Δt, which is the time between first and second laser pulse. The final enhancement in emission (E) under double pulse excitation has been obtained by normalizing the sum of plasma emission under single ($I_1$) and dual pulse LIBS ($I_2$) by emission from single pulse LIBS [20]. During this normalization some of the common factors affecting the plasma emission are automatically canceled out.

$$E = 1 + \left(\frac{\dot{m}_2}{\dot{m}_1}\right)^2 \frac{\left[1 - \exp\left\{-\frac{32}{15} \frac{C_{s1} \Delta t \nu_{ei}}{c}\right\}\right]}{\left(\frac{\Delta t}{T_1} + \frac{C_{s2}}{C_{s1}} \frac{T_2}{T_1}\right)} \quad (6)$$

Where $\frac{\dot{m}_2}{\dot{m}_1}$ and $C_{s2}/C_{s1}$ is the ratio of mass ablation rate and the plasma expansion speed after double and single pulse LIBS respectively. $\nu_{ei}$ is the electron ion collision frequency, $T_1$ and $T_2$ is the time delay for recording the LIBS signal after laser pulse in single pulse LIBS and after second pulse in double pulse LIBS respectively. Here one can take the value of $T_1$ as any value between Δt (for maximum emission) and the time of peak emission in double pulse LIBS (from first laser pulse) without any significant change in the result. Eq.- 6 has been used to study the effect of different parameters on the enhancement in plasma emission under double pulse LIBS, which explained various experimental observations of dual pulse LIBS as reported earlier [20]. According to eq.-6 zero delay between lasers (Δt = 0) provides an enhancement of unity (E =1), just like a single pulse LIBS emission.

**3.2    Emission from Single Pulse LIBS**

The above eq.-6 obtained for dual pulse excitation of the plasma has been simplified for the study of single pulse LIBS under the assumption that the plasma is formed and starts expanding, when initial portion (Pedestal) of the laser pulse interacts with the matter. In this situation rest of the laser energy gets absorbed in the preformed plasma. Therefore, a single laser pulse can be considered operating as dual pulse with certain assumptions. Here one can consider that after the plasma formation by pedestal of the laser, it expands for the period $\tau_L$ (duration of laser pulse) during which laser gets absorbed in the expanding plasma. This indicates that Δt delay between two lasers in eq.-6 can be replaced by $\tau_L$ for the single pulse case. Here eq.-6 can be simplified in two ways as described below to better understand the dependence of important plasma parameters on the emission characteristics of the plasma as well as on its temporal behavior [21].

**Case - 1**

In this case assumption is made that $\Delta t = \tau_L$ (laser time duration), $T_1 = \tau_L/2$ (time of peak laser intensity). $T_2$ is the time at which measurement of the LIBS emission is taking place. Other consideration is that $C_{s2} = 2\, C_{s1}$, that is, initial plasma expansion speed is half of the expansion speed obtained after full laser pulse. In this situation eq.-6 takes the shape as [21]

$$E = 1 + \frac{\left(\dfrac{\dot{m}_2}{\dot{m}_1}\right)^2 \left[1 - \exp\left\{-\dfrac{32}{15}\dfrac{C_{s1}\tau_L V_{ei}}{c}\right\}\right]}{\left(2 + 4\dfrac{T_2}{\tau_L}\right)} \qquad (7)$$

This equation provides information about the emission from the single pulse LIBS at different time delay that means the variation of emission with the gate delay ($T_2$).

**Case – 2**

Here we consider $\Delta t = \tau_L$ and $T_1 = T_2 = T$ is the gate delay or the time of measurement of the emission after the laser peak. Ratio of the laser pulse time duration $\tau_L$ and gate delay T (time of measurement) will be negligibly small for nanosecond time duration laser ($\tau_L / T \approx 0$). After these assumptions eq.-6 is modified as [21]

$$E = 1 + \left(\frac{\dot{m}_2}{\dot{m}_1}\right)^2 \left(\frac{C_{s1}}{C_{s2}}\right)\left[1 - \exp\left\{-\frac{32}{15}\frac{C_{s1}\tau_L v_{ei}}{c}\right\}\right] \qquad (8)$$

This indicates that emission from single pulse LIBS is mainly dependent on the square of the ratio of the mass ablation rate, the ratio of plasma sound velocity, the plasma collision frequency and pulse duration of the laser.

## 4. RESULTS AND DISCUSSION
### 4.1 Effect of Laser Energy
#### 4.1.1 Plasma from Solid Target

The effect of magnetic field on the LIBS spectrum of the aluminum alloy has been reported in the spectral range 350-370 nm, in order to find the emission characteristics of the minor species present in the alloy. The LIBS spectrum is found containing emission lines from various minor elements present in alloy, such as Mn, Fe, Cr and Ti. Similar results have also been reported by many authors for different types of samples [27-29]. LIBS spectrum recorded in the presence of ~5 kG magnetic field shows an enhancement in the intensity of emission lines by nearly a factor of two compared to the spectrum recorded without the magnetic field [23]. Manganese and titanium with very feeble line emission intensity also show similar enhancement in the presence of a magnetic field. It is noticed that the enhancement factor slightly changes from element to element. This enhancement in intensity indicates that the presence of a magnetic field affects the plasma emission characteristics significantly. This enhancement seems to be due to confinement of laser-produced plasma in the presence of a magnetic field.

The effect of laser energy on the LIBS emission can be seen in the LIBS spectra of aluminum alloy sample recorded in the absence as well as in the presence of magnetic field, where energy of the laser was varied from 10-150 mJ ($4\times10^9$-$60\times10^9$ W/cm$^2$). The variation of chromium line intensity ($\lambda$=357.87 nm) with laser energy in the absence of a magnetic field (Fig.-1) shows that emission intensity increases with an increase in laser energy up to nearly 20 mJ. Any further increase in laser energy shows saturation in the chromium line emission intensity. Initially the laser radiation is absorbed up to 20 mJ and ablation of the material increases linearly with the laser intensity. It seems that excess laser energy beyond 20 mJ ($\geq$ $8\times10^9$ W/cm$^2$) is not being coupled to laser induced plasma. This is possible only due to shielding produced by the plasma near critical density surface (formed at high laser intensity) or due to saturation in emission. Similar behavior of plasma emission has been reported by Aguilera et al [30]. Probably excess energy is being reflected from the critical density of the plasma, where the plasma frequency becomes equal to the laser frequency and laser can not propagate further toward the higher plasma density side. Self-absorption of emission is also possible due to higher plasma density, when experiment is performed at higher laser intensity. The presence of self-absorption can be observed from the measurement of full width at half maximum (FWHM) of the Lorenzian profile of line emission [30]. Both the processes (shielding and self-absorption) are possible in the plasma, which can contribute towards saturation in the emission. Presence of magnetic field shows no significant effect on atomic line emission intensity below the laser energy of 10 mJ (Fig.-1). An enhancement in the emission intensity in the presence of a magnetic field is noted for laser energy between 10 to 50 mJ. This enhancement in emission intensity decreases, when the laser energy was further increased beyond 50 mJ. The plasma emission in the presence of a magnetic field for laser energy > 80 mJ ($32\times10^9$ W/cm$^2$) was even lower than the emission observed in the absence of a magnetic field. This indicates that along with reflection of laser energy from critical density level and self-absorption of emission, some other loss factors are also added up in the presence of magnetic field. This extra factor may be the generation of instability (density fluctuation) in the plasma in the presence of a magnetic field [31]. Density fluctuations are expected in this situation due to bouncing of the plasma near $\beta = 1$ location (bouncing radius $R_B$), where plasma kinetic energy becomes equal to the magnetic energy. Even some high-frequency instability such as Large Larmor radius instability is also expected in such a plasma condition. The possibility of density fluctuation in the plasma as a result of bouncing of plasma at $\beta = 1$ location and generation of high frequency instability in the presence of similar external magnetic fields have already been discussed and reported [4, 31]. The presence of instability in the plasma will lead to the scattering of laser light and the loss of plasma particles as a result of cross field diffusion. This may be the reason behind the decrease in emission intensity in the presence of magnetic field for the laser energy greater than 80 mJ.

Similar variation in iron line emission at λ= 358.12 nm with laser energy has also been observed in the absence and the presence of magnetic field. The enhancement in the emission intensity for iron ranges between laser pulse energy of 10 - 40 mJ, which is less than the case for chromium. Better energy span in the case of chromium may be due to its higher transition probability in comparison to iron.

**4.1.2 LIBS from the Liquid Target**

LIBS spectra from a liquid jet target having manganese (Mn) in low concentration has been recorded in the absence and presence of magnetic field after 10-µs time delay from the laser pulse so as to avoid the high background emission from the hot plasma (Bremsstrahlung emission). Three strong peaks are observed at the wavelengths of 403.08, 403.31 and 403.45 nm, which are neutral line emissions from the Mn atom. Similar spectrum is observed in the presence of the magnetic field, but with an enhanced line emission by ~ 1.5 times [26]. Enhancement remains nearly same for all the three wavelengths. It is noticed that the intensity of all the three Mn lines increase linearly with the laser energy up to ~ 300 mJ in the absence of magnetic field. But the presence of the magnetic field shows an enhancement in the LIBS intensity up to 200 mJ followed by a saturation (decrease) towards higher laser energy (Fig.-2). The maximum signal enhancement is observed between 150 - 200 mJ of laser energy. Similar enhancement in the emission intensity has been noted in the presence of magnetic field for other elements also such as magnesium, chromium and titanium in aqueous solution. Enhancement factor is found dependent on the transition probabilities of the emission lines and the dynamics of atoms in the plasma plume. The Mn atomic line emission lasts for nearly 40 - 45 µs, whether magnetic field is present or not. The emission intensity remains high in the presence of the magnetic field for all the gate delay between 5 - 30 µs [26]. Either no or a small line emission is noted below 5 µs gate delay mainly due to dominant background emission, which indicates that the magnetic field does not affect emission from the hot plasma.

Although the threshold laser energy needed for the breakdown is ~10 mJ for solid sample in comparison to ~ 50 mJ for a liquid sample, the basic observations remains similar for solid or liquid samples. Finally enhancement in the emission intensity in the presence of magnetic field is noted mainly for moderate laser energy after the breakdown.

**4.2 Temporal Evolution of LIBS Emission**

The temporal evolution of emission intensity for chromium line from aluminum alloy in the absence and the presence of a magnetic field (Fig.-3) shows that the emission increases as the gate delay increases from 1 μs to 2 μs after which it started decreasing. The emission intensity lasts for nearly 20 μs. An over all increase in intensity is noted for all the delay in the presence of magnetic field, whereas shape of the curve remains same. This shows that initially plasma remains hot and emission from it is dominated by the background emission mainly due to Bremsstrahlung emission. Normally plasma expands away from the target and gets cool with an increase in gate delay, which results in an increase in the probability of radiative recombination. Finally the presence of magnetic field confines the laser produced plasma and increases the effective density of plasma in the emission volume, which ultimately increases the probability of radiative recombination.

The experimental observations are compared with the simple model presented as eq.-3 and 5. It has been shown that enhancement in plasma emission (Eq.5) and deceleration in the plasma expansion (Eq.3) change with plasma β as shown in Fig.-4. The value of plasma β depends mainly on the density and temperature of the plasma, where both decreases as the plasma expand away from the target. It also depends on the intensity of magnetic field, which remains constant during the experiment. For comparison with theory, one can consider $t_1 = t_2$ in eq. (5), because the total emission lasts for nearly the same time (~20 μs) in the absence as well as in the presence of magnetic field (Fig.-3). Fig.-4 clearly shows that initially for higher value of β, plasma deceleration and enhancement in emission are negligible. However as the β decreases below 10, the ratio of plasma expansion velocities $v_2/v_1$ decreases and the ratio of emission intensity $I_2/I_1$ (value of enhancement) increases. This indicates that Bremsstrahlung emission cannot contribute to the enhancement in the emission, because it occurs at higher plasma temperatures and density (β > 10) at comparatively early time (smaller gate delay) after the plasma formation. Only line emissions and recombination radiations can contribute to the enhancement at decreased plasma temperature and density during plasma expansion. This is in good agreement with the experimental observations.

An increase in effective density of emitting plasma due to magnetic confinement has been verified by recording the stark broadening in the $H_\alpha$ emission. $H_\alpha$ spectra are recorded from the aqueous solution of magnesium around 656 nm in the absence and presence of magnetic field, which shows Stark broadening in the spectral line emission. This indicates toward an increase in plasma density in the presence of magnetic field due to deceleration of plasma [23] as a result of magnetic confinement. Temporal evolution of plasma density is inferred by measuring Stark broadening with change in gate delay in the absence and presence of magnetic field. Plasma density is found changing from $10^{19}$ to $10^{17}$ particles/cm$^3$ with a change in delay (0.5 – 10 µs). No significant change in plasma density is noted for gate delay of 2 – 3 µs. Plasma density shows an increase beyond 4 µs, which indicates that confinement becomes effective only after sufficient plasma expansion when plasma density and temperature decreases sufficiently resulting in low value of plasma β. The decrease in plasma temperature due to plasma expansion and an increase in effective plasma density due to magnetic confinement increase the probability of radiative recombination, which seems to be the reason for an increase in plasma emission. This is in agreement with the results obtained from the simple model presented here, which is successfully explaining the experimental observations. Finally Eq. (5) and Fig.-4 shows that in principle emission can be increased even more than ~2 times by limiting the effective value of plasma β close to one. This is possible only by keeping the laser energy moderate as well as by increasing the strength of the steady magnetic field. Further investigation in this direction with an increased magnetic field can throw more light on this subject.

### 3.3    Saturation in Plasma Emission

The saturation in plasma emission during LIBS has been observed mainly at higher laser intensity. The intensity of laser mainly decides the density as well as the temperature of the plasma. The external magnetic field also affects the saturation threshold in the plasma emission as a result of magnetic confinement. The study of variations in plasma emission with electron density and temperature for laser of different time duration can provide better understanding of saturation in the plasma emission. For this purpose calculation is performed using eq.-8 with a consideration that $\dfrac{\dot{m}_2}{\dot{m}_1} = 4$, $\dfrac{C_{s2}}{C_{s1}} = 2$ and $T_e = 1\text{eV}$ and the results are presented in Fig.-5, which shows variation in plasma emission with change in plasma density for different time duration of the laser $\tau_L$. It shows that the plasma emission starts increasing for plasma density of ~ $10^{16}$ cm$^{-3}$ for a laser of 1 ns time duration. This increase becomes fast with an increase in plasma density and saturates at plasma density of~ $10^{19}$ cm$^{-3}$. Similar variations are noted for 10 and 20 ns time duration laser, but with minor change in the threshold for initiation and saturation of emission, which occur at lower density for large time duration lasers. Higher time duration laser produce comparatively more plasma emission as a result of more material ablation [32-33]. In this case the electron-ion collision frequency and laser plasma interaction time play an important role because plasma collision frequency increases with an increase in the plasma density. It helps in increasing the absorption of laser in plasma through inverse Bremsstrahlung and consequently the plasma emission, whereas large time duration of laser provides longer time for plasma to expand and absorb the laser efficiently.

The effect of plasma temperature on the emission has also been estimated using eq.-8 where plasma density is kept constant at $\sim 10^{17}$ cm$^{-3}$ and all the other parameters remains same as in Fig.-5. In this case emission from plasma is calculated by varying the plasma temperature induced by different time duration laser. Fig.-6 shows that plasma emission is very high for low temperature plasma. It decreases initially very fast and then slowly with an increase in plasma temperature. The decrease in plasma emission with an increase in plasma temperature is due to decrease in electron-ion collision frequency as $\nu_{ei} \propto (T_e)^{-3/2}$. Here again plasma emission is high for longer time duration laser as seen in Fig.-5. In this case saturation occurs at lower plasma temperature near 1 eV. Results of Fig.- 5 & 6 indicates that saturation in the plasma emission is possible only when the plasma density is high and plasma temperature is low. In other words one can say that instead of plasma density and plasma temperature, electron-ion collision frequency plays an important role in deciding the saturation in plasma emission. The value of electron-ion collision frequency for saturation has been obtained as $\nu_{ei} > 10^{13}$ s$^{-1}$.

## 4.   EFFECT OF PLASMA CONFINEMENT

It is well known that laser produced plasma expands away very fast from the target surface and emits various types of emission ranging from X-ray to visible emission depending on the temperature of the plasma. Both types of plasma emission increase after plasma confinement [21]. Two types of plasma confinement have been reported as magnetic and spatial confinement for enhancing the emission from laser produced plasma. In both the cases plasma expansion is decelerated. The effect of variation in expansion velocity on the plasma emission has been obtained using eq.-8. This calculation has been performed for electron-ion collision frequency of $\nu_{ei} \sim 3.62 \times 10^{12}$ s$^{-1}$ corresponding to plasma density of $10^{17}$ cm$^{-3}$ and plasma temperature of 1eV. The value of $C_{s2}/C_{s1}$ has been changed from 0.1 to 10. Fig.-7 shows that emission intensity is very high when the value of $C_{s2}/C_{s1}$ is low towards 0.1 (deceleration), whereas it decreases very fast as the value of $C_{s2}$ increases. This observation shows that deceleration of expanding plasma plume enhances the plasma emission. It seems that emitting plasma volume increases with an increase in $C_{s2}$ where corresponding energy density of the absorbed laser decreases, which is reflected as decreased plasma emission. This indicates that enhancement in emission is mainly dependent on the order of plasma confinement (deceleration of plasma plume) [21]. This supports the results of enhancement in emission by magnetic and spatial confinement as reported earlier [25]. More enhancement in spatial confinement is due to comparatively better confinement/deceleration of the plasma. In the case of magnetic confinement plasma diffuses across the magnetic field as a result of plasma instability and fluctuations in the plasma, which degrades the plasma confinement, where as spatial confinement is free from these problems. However combination of both the confinement technique is beneficial as has been reported by Guo et al. [25]. Enhancement in plasma emission is independent of the technique used for plasma confinement. Only one factor seems to be important, that is efficiency of the plasma confinement. This result further indicates that confinement of dual pulse produced plasma may provide even more enhancement in emission than the combination of magnetic and spatial confinement.

### 4.1 Temporal Evolution of Emission from Confined Plasma

The temporal evolution of plasma emission can be obtained using eq.-7. For this purpose it is assumed that $T_1 = \tau_L/2$ (Peak time of laser) and $T_2$ is time delay at which the emission intensity is recorded. The effect of plasma confinement and the laser time duration on the temporal evolution have been obtained by calculating emission, (1) for different values of $C_{s2}/C_{s1}$ as 2, 0.5 and 0.1 keeping laser time duration constant at 10 ns, (2) for 1, 10 and 20 ns time duration laser keeping $C_{s2}/C_{s1}$ constant at ~ 2. Fig.-8 shows the variation of plasma emission with time delay for decreasing value of $C_{s2}/C_{s1}$, which indicates effect of plasma confinement on emission. In the case of $C_{s2}/C_{s1}$ ~ 2 plasma emission first decreases very fast and then slowly with time delay. After addition of confinement effect by decreasing the values of $C_{s2}/C_{s1}$ to 0.5 and 0.1, plasma emission increases for all the delay time. Even the decay rate of emission intensity changes. This indicates that plasma emission last for a longer time in the case of plasma confinement. The maximum enhancement in the emission after confinement has been found for time delay of 20 ns. This happens because of slowing down of decay in plasma emission as a result of plasma confinement. Fig.-9 shows variation in temporal evolution of plasma emission obtained from laser having time duration of 1, 10 and 20 ns. Here plasma emission is less for the plasma induced by 1 ns laser, which initially decay fast and then slowly. The rate of decay of plasma emission decreases with an increase in time duration of the laser. Even emission is higher for the plasma produced by longer time duration laser. This may be due to efficient heating of the plasma for longer time. It also shows that plasma produced by long time duration laser last for longer time.

Similar observation has been reported experimentally [23]. They have studied temporal evolution of chromium line emission from aluminum alloy in the absence and presence of a magnetic field (magnetic confinement) and reported that plasma emission was maximum at ~ 2 μs after which it started decreasing. The emission intensity lasted for nearly 20 μs. An overall increase in intensity was noted for all the delay in the presence of magnetic field (magnetic confinement), whereas shape of the curve remained same. This result is similar as shown in Fig.-8. The decay in emission with time indicates that initially plasma is hot and emission from it is dominated by the background emission, which contains mainly Bremsstrahlung emission. As the gate delay increases, plasma expands away from the target and gets cool resulting a decrease in emission. The presence of magnetic field increases the effective density of plasma in the emission volume as a result of confinement of laser produced plasma, which ultimately increases the probability of radiative recombination, electron-ion collision frequency as well as absorbed energy per unit volume. All of these factors play role in increasing the plasma emission. The enhancement in plasma emission during magnetic confinement has also been found correlated with deceleration of the plasma expansion as discussed in previous sections. Mainly line emissions are enhanced due to recombination radiation or heating of the plasma during confinement/deceleration of plasma plume. Many experimental observations reported earlier are in qualitative agreement with these theoretical finding [32-33].

## 6. CONCLUSIONS

The emission intensity in LIBS has been found dependent on the nature of target material (solid or liquid), the transition probability of the elements, the laser intensity and its time duration as well as on the dynamics of plasma plume. LIBS intensity increases linearly in the moderate intensity range of laser whereas higher laser intensity generates saturation in LIBS signal. The deceleration in expansion of plasma plume in the presence of magnetic field enhances the emission from the plasma. The threshold of saturation in plasma emission decreases to lower laser intensity under the effect of magnetic confinement of plasma. Theoretical analysis also supports the experimental findings that enhancement in emission is correlated with the deceleration in plasma expansion. No enhancement in plasma emission was observed, when the plasma β was high (>10) due to high plasma temperature and density. The plasma emission in the presence of a confinement has been found independent of the wavelength of emission and the techniques of confinement as magnetic or spatial. Theoretical analysis of the laser induced plasma shows clearly that saturation in the LIBS emission is mainly dependent on the electron ion collision frequency of the plasma. It is mainly dependent on the density and temperature of the plasma, which plays an important role in the efficient absorption of the laser in the preformed plasma followed by enhanced emission. Finally the time duration of the laser has also been found effective in increasing the LIBS intensity. This study indicates that proper technical arrangement (combination of techniques) of plasma deceleration/confinement may be helpful in increasing the LIBS intensity manifold.

**FIGURE CAPTIONS**

1. Variation in the line emission intensity of chromium ($\lambda = 357.87$ nm) at 5 μs gate delay and 5 μs gate width in the spectra of aluminum alloy with change in laser energy in the absence (B = 0 kG) and presence of (B = 5 kG) magnetic field.

2. Variation in the emission intensity of manganese ($\lambda = 403.07$ nm) in liquid sample with change in laser energy for gate delay and gate width of 10 μs (a) in the absence (B = 0 kG) and (b) presence of magnetic field (B = 5 kG).

3. Variation in the intensity of chromium line emission ($\lambda = 357.87$ nm) from LIBS of aluminum alloy with change in gate delay at 14 mJ laser pulse energy in the absence (B = 0 kG) and presence of magnetic field (B = 5 kG). Gate width is 5 μs.

4. Variation in the ratio of emission intensity ($I_2/I_1$) from plasma and the ratio of plasma expansion velocity ($v_2/v_1$) with change in plasma β. The values of ($v_2/v_1$) and ($I_2/I_1$) were obtained from Eq. (3) and Eq. (5) keeping $t_1 = t_2$, because no change in total emission time was noted for emission in the absence and presence of magnetic field.

5. Variation in emission intensity with an increase in plasma density produced by laser having time duration 1, 10 and 20 ns.

6. Variation in emission intensity with an increase in plasma temperature for plasma produced by 1, 10 and 20 ns time duration.

7. Variation in emission intensity with an increase in ratio of expansion velocity ($C_{s2}/C_{s1}$) for plasma induced by 1, 10 and 20 ns time duration laser.

8. Variation in emission intensity with time delay from the peak of laser for ratio of plasma expansion velocity 0.1, 0.5 and 2.

9. Variation in emission intensity with time delay after laser peak for plasma induced by 1, 5 and 10 ns time duration.

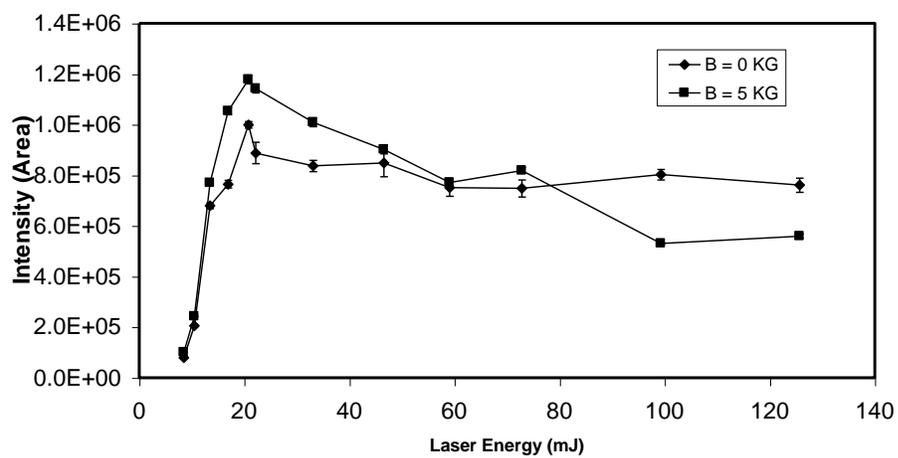

**Fig.- 1**

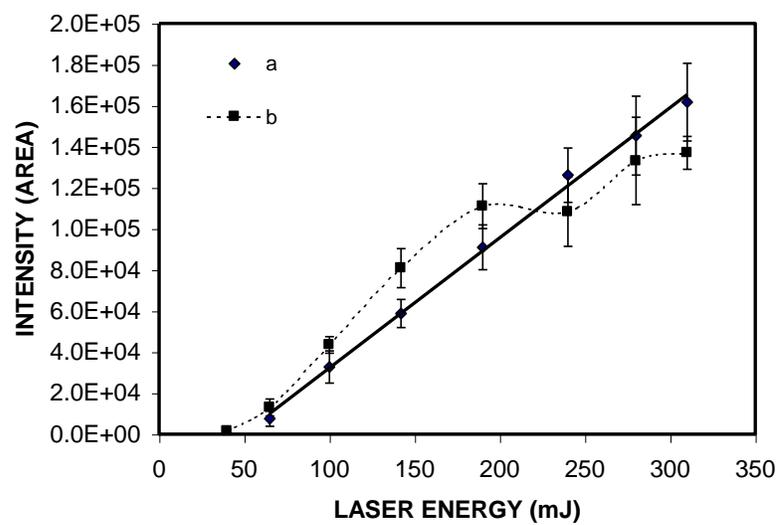

**Fig.-2**

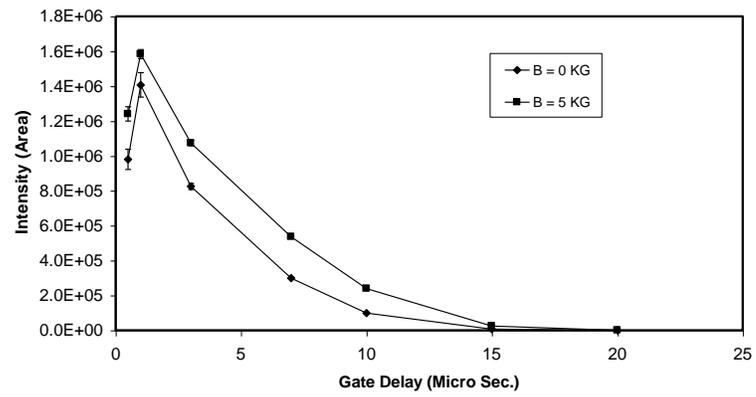

**Fig.-3**

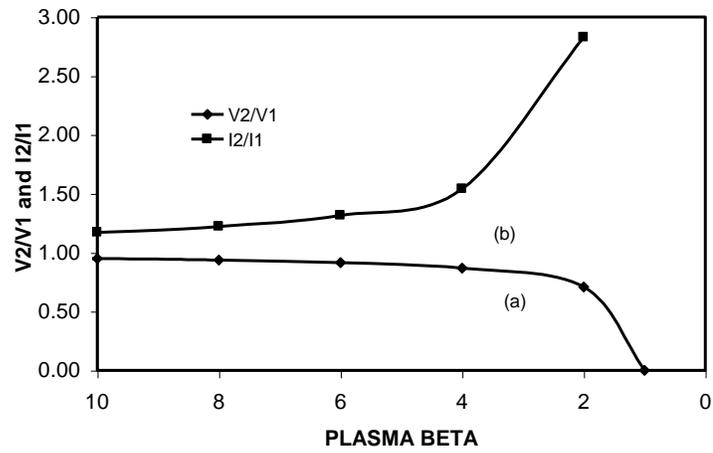

**Fig.-4**

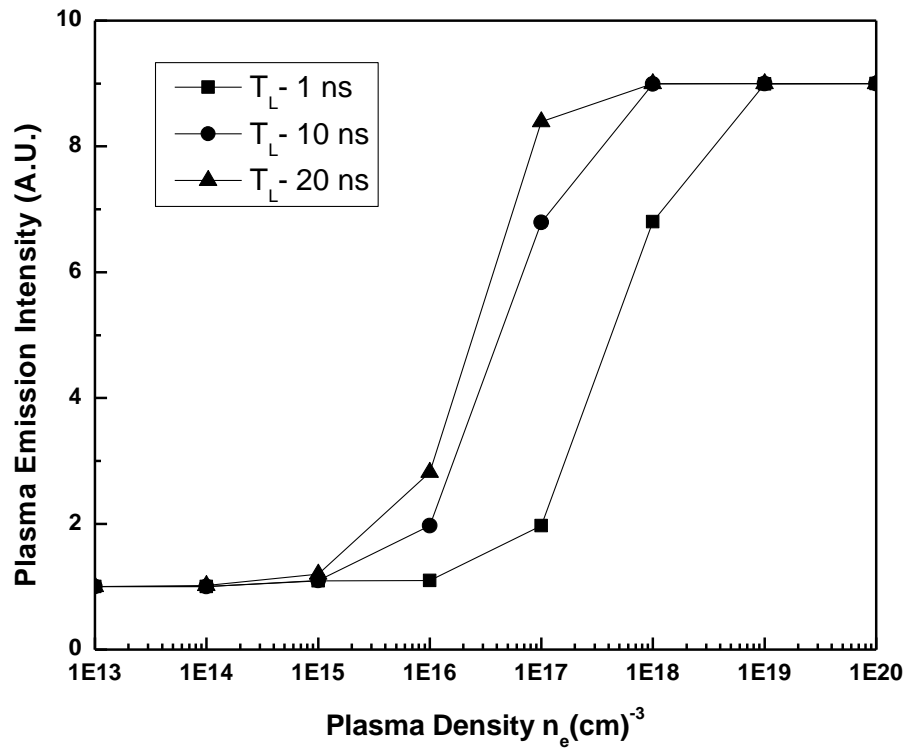

**Fig.-5**

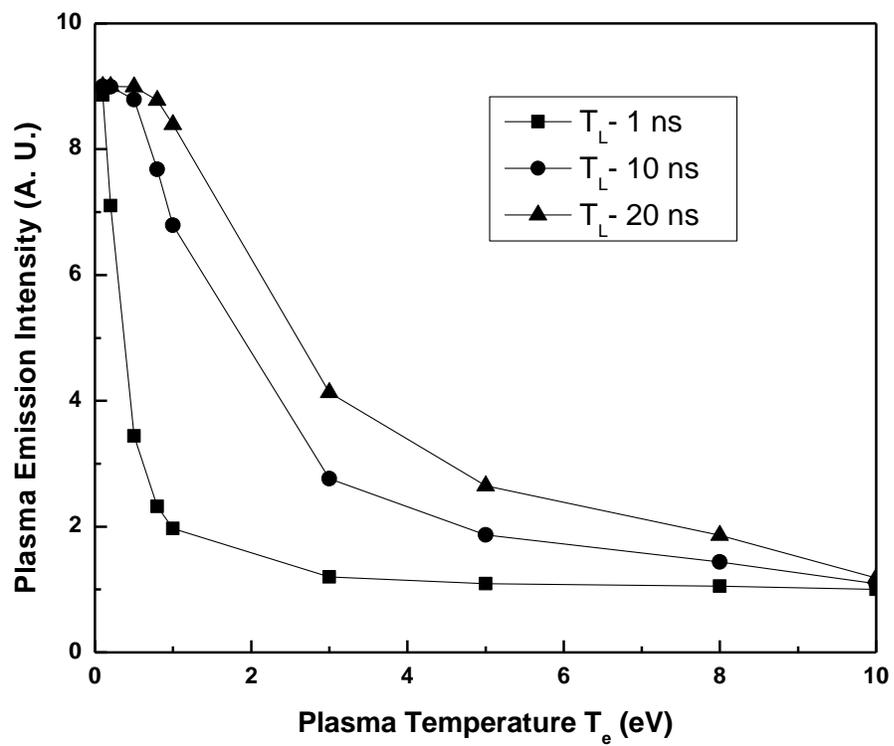

**Fig.-6**

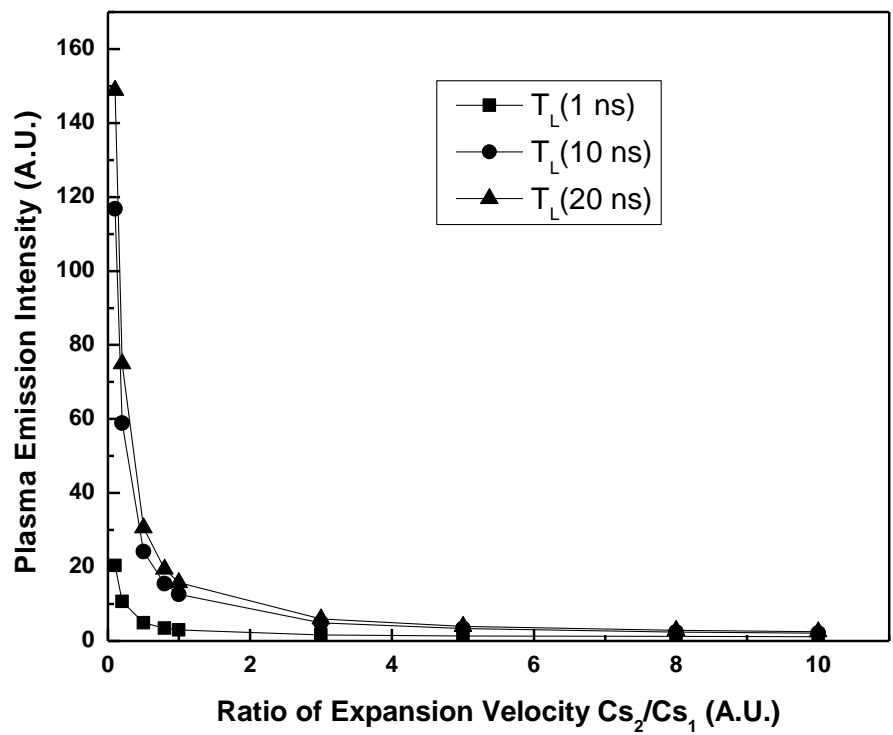

**Fig.-7**

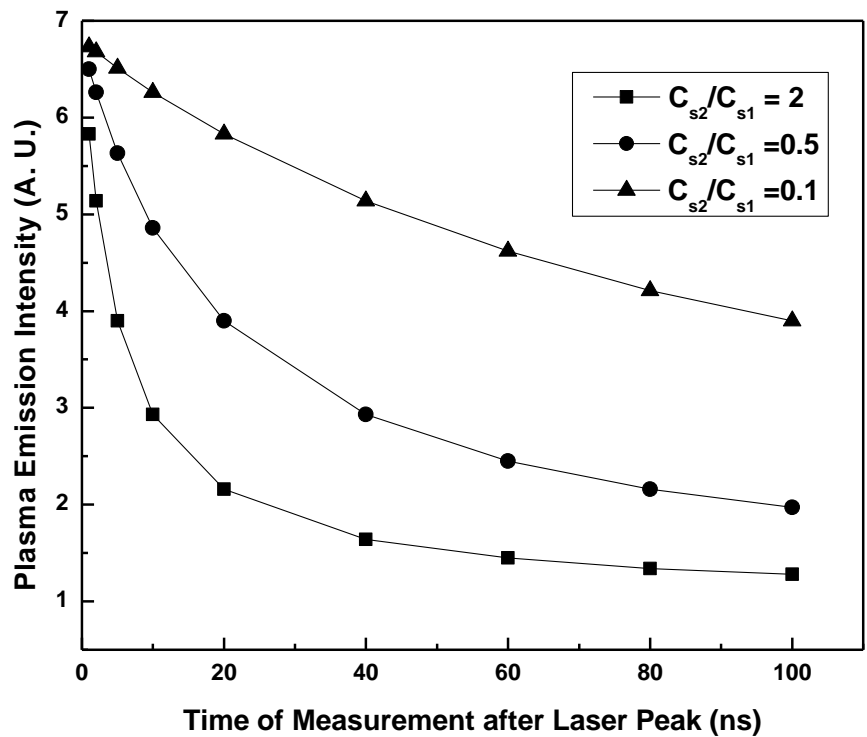

**Fig.-8**

25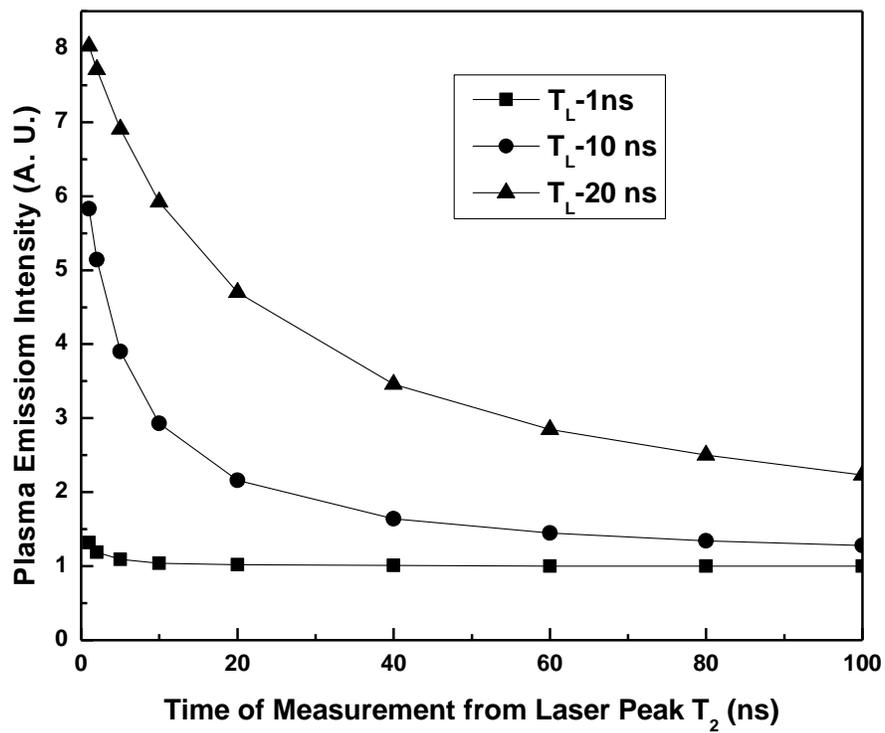

**Fig.-9**